\documentclass[twocolumn, twocolappendix]{aastex7}

\usepackage{bm}
\usepackage{amsmath,amssymb,amsfonts}%
\usepackage{hyperref}

\newcommand{\dg}{$^{\circ}$}

\newcommand{\arcspix}{$^{\prime \prime}~{\rm pix^{-1}}$}
\newcommand{\rsun}{R$_\odot$}

\newcommand{\alfven}{Alfv\'en}

\received{2025 April 05}
\revised{2025 April 16}
\accepted{2025 September 02}
\shorttitle{}
\shortauthors{Shi et al.}

\begin{document}

\title{The Role of Far-side Magnetic Structures in Modeling 2024 Solar Eclipse}

\author[0000-0001-7397-455X]{Guanglu Shi}
\affiliation{Key Laboratory of Dark Matter and Space Astronomy, Purple Mountain Observatory, 
\\ Chinese Academy of Sciences, Nanjing 210023, People's Republic of China}
\affiliation{School of Astronomy and Space Science, University of Science and Technology of China, 
\\ Hefei, Anhui 230026, People's Republic of China}
\email{shigl@pmo.ac.cn}

\author[0009-0001-4778-5162]{Jiahui Shan}
\affiliation{Key Laboratory of Dark Matter and Space Astronomy, Purple Mountain Observatory, 
\\ Chinese Academy of Sciences, Nanjing 210023, People's Republic of China}
\affiliation{School of Astronomy and Space Science, University of Science and Technology of China, 
\\ Hefei, Anhui 230026, People's Republic of China}
\email{shanjh@pmo.ac.cn}

\author[0000-0003-4655-6939]{Li Feng}
\affiliation{Key Laboratory of Dark Matter and Space Astronomy, Purple Mountain Observatory, 
\\ Chinese Academy of Sciences, Nanjing 210023, People's Republic of China}
\affiliation{School of Astronomy and Space Science, University of Science and Technology of China, 
\\ Hefei, Anhui 230026, People's Republic of China}
\email[show]{lfeng@pmo.ac.cn}

\author[0000-0003-3060-0480]{Jun Chen}
\affiliation{Key Laboratory of Dark Matter and Space Astronomy, Purple Mountain Observatory, 
\\ Chinese Academy of Sciences, Nanjing 210023, People's Republic of China}
\email{chenjun@pmo.ac.cn}

\author[0000-0001-9979-4178]{Weiqun Gan}
\affiliation{Key Laboratory of Dark Matter and Space Astronomy, Purple Mountain Observatory, 
\\ Chinese Academy of Sciences, Nanjing 210023, People's Republic of China}
\affiliation{University of Chinese Academy of Sciences, Nanjing 211135, People's Republic of China}
\email{wqgan@pmo.ac.cn}

\begin{abstract}

The corona is a crucial region that connects the solar surface to the solar wind and serves as the primary site of solar activity. The 2024 total solar eclipse (TSE) provides a unique opportunity to investigate the large-scale coronal structure. Combined with TSE observations, we study the impact of the magnetic structure of the far-side active region, located in the eastern hemisphere of the Sun that has not yet rotated into the Earth Field-of-View (FoV), on a global Magnetohydrodynamic (MHD) simulation. To address the limitation of single-view measurements in the routine synoptic map, we correct the magnetic field in the far-side region by incorporating full-disk magnetograms measured several days after the TSE, allowing us to capture the temporal evolution of the photospheric magnetic field in near real-time. Simulation results demonstrate that the local magnetic field in the far-side active region can significantly influence the global coronal structure by altering the position of the heliospheric current sheet (HCS), and further affect the global distribution of plasma parameters, even in polar regions. A comparison of the simulation results with white-light (WL) TSE + LASCO C2 observations and in situ measurements by the Parker Solar Probe (PSP) reveals that the composite synoptic map improves the accuracy of coronal modeling. This work provides robust support for advancing our understanding of coronal evolution, as well as deepens the link between the photosphere and large-scale coronal structure. Furthermore, it establishes a theoretical foundation for the future development of multi-view, stereoscopic measurements of the photospheric magnetic field.

\end{abstract}



\section{Introduction} \label{sec1}

The corona is the outermost layer of the solar atmosphere, characterized by its complex and dynamic structure, closely associated with solar activity \citep{Aschwande2005}. Solar eruptive events are the primary source of catastrophic space weather, with CME-driven shocks capable of accelerating particles to extremely high energies, posing significant threats to Earth's ecosystem and the Sun-Earth space environment. Investigating the corona is not only fundamental to advancing our understanding of solar physics, but also critical for improving space weather predictions and assessing their geomagnetic effects.

Due to the high temperature and low density of the corona, directly observing its fine structures and dynamic evolution remains a great challenge. Consequently, an essential approach to studying the corona and improving space weather forecasting is through physical modeling of the solar atmosphere and interplanetary space. This objective achieved by coupling a series of models that self-consistently connect the solar surface to the near-Earth environment, such as the SWMF \citep{Toth2012}, WSA-ENLIL \citep{Wang1990, Arge2000, Odstrcil2005}, EUHFORIA \citep{Pomoell2018}, MAS \citep{Linker1999}, COCONUT \citep{Perri2022ApJ}, SWASTi \citep{Mayank2022}, and other large-scale Magnetohydrodynamic (MHD) numerical simulation frameworks. With the increasing availability of observations provided by advanced solar instruments \citep{Pesnell2012SoPh, Liu2014, Fox2016, Rimmele2020, Muller2020, Li2019165, Gan2023SoPh}, these MHD models have been extensively developed, improved, and validated. 

The total solar eclipse (TSE) provides a unique opportunity to observe and study the corona, as well as to validate and refine model predictions \citep{Wagner2022, Benavitz2024}. When the moon precisely aligns between the Sun and Earth, it completely obscures the photospheric light, allowing the intricate structures of the corona to appear. Numerous studies \citep{Mikic2018, Boe2021, Boe2022, Boe2023, Molnar2025, Young2025ApJ, Liu2025, Downs2025} have employed TSE observations in combination with simulations to advance our understanding of coronal fine structures, heating mechanisms, brightness distributions, and related phenomena. In particular, polarized brightness (pB) measurements obtained during TSEs enable the inference of further coronal plasma parameter distributions, thereby providing deeper insight into the physical processes occurring in the corona \citep{Hanaoka2021, Hanaoka2024, Bemporad2023ApJ}.

The computational accuracy of MHD models largely depends on the precision and timeliness of the input photospheric magnetic field. As the source of the magnetic field in the solar atmosphere, the photospheric magnetic field is typically measured by ground- or space-based instruments \citep{Harvey1996, Scherrer2012, Deng2019RAA}. Notably, polar magnetic fields have been demonstrated to play a significant role on simulations of the corona and interplanetary space \citep{Riley2019, Shi2024}. However, observational limitations, particularly the lack of direct measurements of the solar far-side and polar regions, necessitate the construction of synoptic maps by splicing together magnetograms observed from Earth view at different times. While this approach fills the spatial and temporal observational gaps to a certain extent, it also introduces potential uncertainties.

To address these issues, this work conducts an MHD simulation of the TSE that occurred on 2024 April 8 over North America. By comparing simulation results derived from different synoptic maps, we aim to explore potential ways to improve the accuracy of coronal modeling. Section \ref{sec2} describes the observational and simulation methodologies used for the 2024 TSE. Section \ref{sec3} presents our results: (1) the distribution of plasma parameters, magnetic topology, and energy; (2) a comparison of in situ measurements from the Parker Solar Probe (PSP) and white-light (WL) observations with simulation outputs. Discussion and conclusions are given in Section \ref{sec4}.

\section{Methodology} \label{sec2}

\subsection{Observations}

The TSE images were captured in Jackman, Maine, USA, starting at 19:30 UT on 2024 April 8, with a total duration of 3 minutes 24 seconds. The observational conditions were nearly perfect, with a completely clear sky and no visible clouds. The primary objective was to obtain high-resolution polarized images of the corona for comparison with MHD simulations. Due to the temporary nature of the field observation, a portable optical system was assembled to efficiently capture data within the limited observing window. The core of the optical system was an Askar V refractor telescope with a focal length of 480 mm and an f-ratio of F/6. A 2-inch Andover broadband filter, centered at 590 nm with a bandwidth of 100 nm, was installed in the filter drawer. This waveband is consistent with the Large Angle and Spectrometric Coronagraph \citep[LASCO;][]{Brueckner1995} onboard the Solar and Heliospheric Observatory \citep[SOHO;][]{Domingo1995}, facilitating the effective detection of scattered coronal light. To analyze the polarization properties of the corona, a filter wheel equipped with three 1-inch Codixx linear polarizers was used to achieve imaging at three polarization angles of $-60${\dg}, 0, and $+60${\dg} to diagnose the coronal electron density distribution and infer the magnetic field structure. The imaging system utilized a ZWO ASI533MM Pro cooled monochrome camera with a Sony IMX533 CMOS sensor, characterized by high quantum efficiency and low readout noise, ensuring a high dynamic range for capturing coronal structures. The sensor has a resolution of 3008$\times$3008 ${\rm pix^{2}}$ with a pixel size of 3.76 ${\rm \mu m}$, and provides a spatial scale of 1.62{\arcspix} paired with the Askar V refractor telescope. A ZWO AM5 equatorial mount was used to maintain stable solar tracking and minimize image drift during long-exposure imaging. The entire observing system was carefully calibrated and optimized to ensure the reliability of the acquired data.

\begin{figure*}[htb!]
\begin{center}
\includegraphics[width=0.8\textwidth]{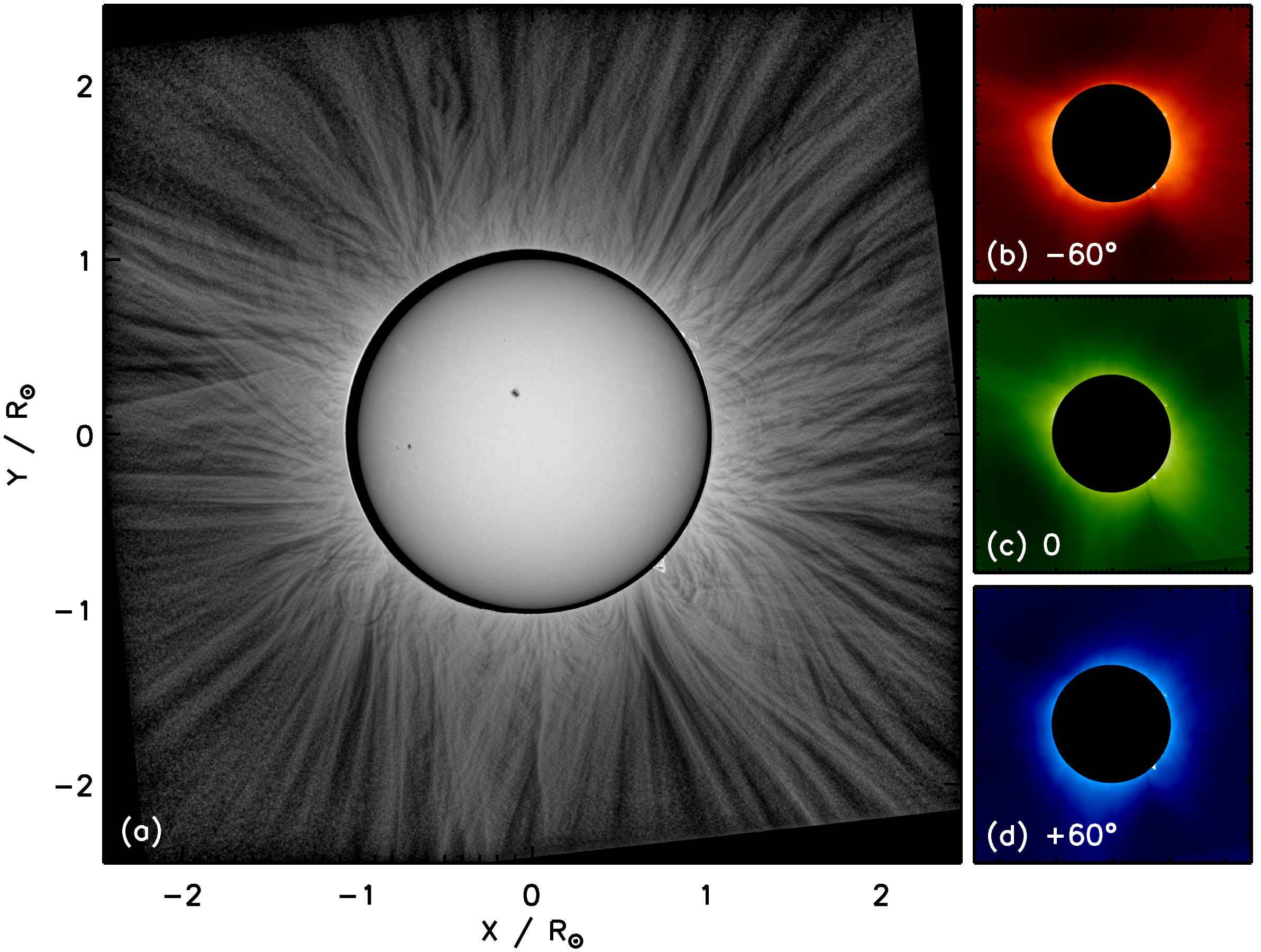}
\caption{WL observations of the 2024 TSE enhanced by the WOW algorithm (left column) and radial filtering method (right column). Panel (a) presents a composite image of the coronal pB and the solar disk with a FoV of 2.45~{\rsun}, where the coronal structures are observed using polarizers oriented at three angles of $-60${\dg} (b), 0 (c), and $+60${\dg} (d), respectively. Each polarized image, shown in different colors, has been corrected for flat-field and dark-field, co-alignment, and exposure time normalization. \label{fig1}}
\end{center}
\end{figure*}

Figure \ref{fig1} presents the WL polarized observations of the 2024 TSE, observed by the instruments described above, with a Field-of-View (FoV) of 2.45~{\rsun}. The wavelet-optimized whitening \citep[WOW;][]{Auchere2023} algorithm (left column) and the radial filtering method (right column) are applied to enhance coronal features, respectively. Panel (a) is a pB image of the corona synthesized from three polarized images, with a WL image of the solar full-disk overplotted, clearly revealing the prominences and large-scale coronal structures. The right column shows the images with the polarization angles of $-60${\dg} (b), 0 (c), and $+60${\dg} (d) after flat-field and dark-field corrections, co-alignment, and exposure time normalization. Each image is a composite of long-exposure (60 ms) and short-exposure (11 ms) observations to enhance its dynamic range. Flat-field and dark-field calibration frames were obtained after the TSE observation by pointing the system at the sky with the zoom lens cover open and closed, respectively. Simultaneously, the full-disk image in the same waveband, as shown in panel (a), was used to calculate the mean solar brightness (MSB), which in turn enabled the radiometric calibration of the three polarized images. Consequently, the pB and total brightness (tB) of the corona during the 2024 TSE were derived \citep{Feng2019}. The attenuator for disk imaging has a nominal attenuation coefficient of $1 \times 10^{-5}$, which may be underestimated due to possible abrasion. Consequently, both pB and tB may be underestimated. These high-resolution observations provide crucial constraints for subsequent numerical simulations and coronal studies.

\subsection{Simulations}

\begin{figure*}[htb!]
\begin{center}
\includegraphics[width=0.85\textwidth]{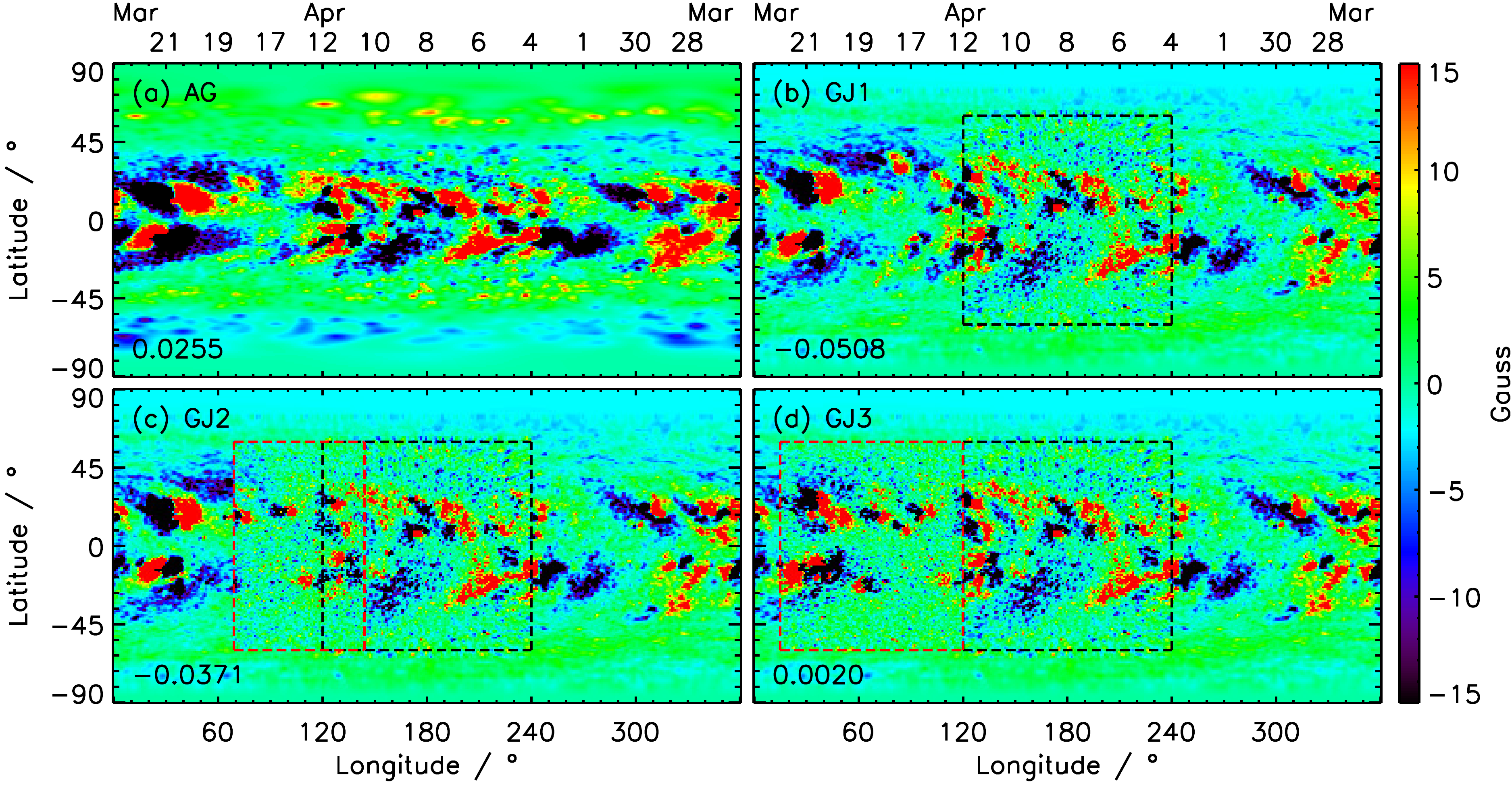}
\caption{Synoptic maps used in the 2024 TSE simulation. The 7th realization of the ADAPT-GONG map (a, labeled as AG) at 18:00 UT and the standard GONG Janus daily synoptic map (b, labeled as GJ1) at 19:04 UT on 2024 April 8. The black dashed box represents the region where a 1-hour averaged full-disk magnetogram is replaced in the original GJ1 map. The composite synoptic maps, labeled as GJ2 (c) and GJ3 (d), generated by replacing the GONG full-disk magnetograms measured after 4 and 8 days on the corresponding region (represented by the red dashed boxes) in the GJ1 map. Each panel includes the magnetic flux balance coefficient. The viewpoint of Earth is shifted to Carrington longitude 180{\dg} for better visualization. The date corresponding to Carrington longitude is indicated at the top of the panels. \label{fig2}}
\end{center}
\end{figure*}

The 2024 TSE is simulated by the Solar Corona (SC) component in the SWMF based on the {\alfven} Wave Solar atmosphere Model-Realtime \citep[AWSoM-R, ][see Appendix \ref{sec:apd1} for more details]{Sokolov2021}. The SC applies an adaptive three-dimensional (3D) spherical stretched grid in the HelioGraphic Rotating (HGR) coordinate, covering the coronal region from 1.05~{\rsun} to 24.0~{\rsun}. The photospheric magnetogram, serving as the bottom boundary condition of the data-driven AWSoM-R model, is critical for accurately simulating the corona and solar wind. Due to the lack of stereoscopic measurements of the photospheric magnetic field, the synoptic map can be typically constructed by single-point measurements obtained from Earth view over an entire Carrington Rotation (CR). To better capture the real-time evolution of active region magnetic fields, a common approach involves updating the synoptic map with the most recent full-disk magnetogram, such as the GONG Janus map. Nevertheless, the only viable solution for measuring the evolution of the magnetic field in the far-side active region, where the part of the Sun that has not yet rotated into the Earth FoV, is to incorporate multi-view observations from the Polarimetric and Helioseismic Imager \citep[PHI;][]{Solanki2020} onboard the Solar Orbiter \citep[SolO;][]{Muller2020}.

To evaluate the impact of far-side active regions on the 2024 TSE simulation, we assume their magnetic structures remain nearly stable in the next 8 days, and then incorporate GONG full-disk magnetograms observed 4 and 8 days after April 8 into the GONG Janus map. Figure \ref{fig2} presents the synoptic maps used as boundary conditions. As a comparison, panel (a) presents the 7th realization of the ADAPT-GONG map (labeled as AG) after correction by the Air Force Data Assimilation Photospheric Flux Transport \citep[ADAPT;][]{Arge2010, Hickmann2015} model, widely used in MHD simulations. Panel (b) presents the standard GONG Janus map (labeled as GJ1), where the black box indicates the region updated with a 1-hour averaged full-disk magnetogram on that day. The AG and GJ1 maps used in this work were generated at 18:00 UT and 19:04 UT, respectively, on 2024 April 8. The top horizontal axis denotes the date corresponding to the Carrington longitude. Panels (c) and (d) present the composite maps (labeled as GJ2 and GJ3) incorporating part of the full-disk magnetograms from 4 and 8 days later, respectively. The red boxes indicate the replaced regions, covering the regions within $-60${\dg} $\sim$ $+15${\dg} and $-60${\dg} $\sim$ $+45${\dg} from the central meridian, respectively, while spanning the latitude from $-60${\dg} to $+60${\dg}. The replacement strategy aims to update the far-side region while ensuring the integrity of the active region and preserving as much of the original data within the black boxes as possible. To facilitate comparison, the viewpoint of Earth is shifted to Carrington longitude 180{\dg}. Each map has dimensions of 360 (longitude, $\phi$) $\times$ 180 (latitude, $\theta$). The magnetic flux balance coefficient ($c = \Phi_B/|\Phi_B|$) is indicated in each panel, where values closer to zero reflect a more balanced distribution of positive and negative flux. It can be found that the AG map exhibits strong positive flux, while the GJ3 map achieves near-perfect flux balance. The magnetic field strength in the active and polar regions of the AG map is higher than that in the GJ1–GJ3 maps, and the two types of maps also exhibit opposite magnetic polarities in the north polar region. These differences directly result in variations in the coronal plasma parameters derived from the TSE simulations.

The initial distribution of the coronal magnetic field is extrapolated by the Potential Field Source Surface \citep[PFSS;][]{Altschuler1969, Schatten1969} model based on the spherical harmonic method, with the sole free parameter, the source surface $R_{\rm ss}$, taken as 25~{\rsun}, and the spherical harmonic order $N_{\rm max}$ set to 180. The source surface is significantly higher than the usual value of 2.5~{\rsun} to eliminate the non-zero curl within the SC domain and minimize numerical artifacts to improve the computational accuracy \citep{Sachdeva2021}.

The initial temperature and number density of electrons and protons at the inner boundary are set to $T_\odot = 5 \times 10^4\ {\rm K}$ and $N_\odot = 2 \times 10^{11}\ {\rm cm^{-3}}$, respectively. The relatively high number density is chosen to effectively avoid the chromospheric evaporation effect without influencing the final simulation results \citep{Lionello2009}. The Poynting flux $S_{\rm A}$ represents the energy density of {\alfven} waves propagating outward from the inner boundary. A key model parameter, the ratio $(S_{\rm A}/B)_\odot$ between $S_{\rm A}$ and the magnetic field $B$, significantly influences plasma parameters and coronal structures. By comparing the simulation results obtained using different values of $(S_{\rm A}/B)_\odot$, we determine that the optimal value is $7 \times 10^5\ {\rm W\,m^{-2}\,T^{-1}}$ for the AG map and $8 \times 10^5\ {\rm W\,m^{-2}\,T^{-1}}$ for the GJ1-GJ3 maps, respectively. The correlation length of the {\alfven} waves, $L_\perp$, scales as $B^{-1/2}$, with the proportionality constant $L_\perp \sqrt{B}$ serving as an adjustable parameter set to $1.5 \times 10^5\ {\rm m\,\sqrt{T}}$. The stochastic heating exponent and amplitude \citep{Chandran2011} related to the energy partitioning between electrons and protons are set to 0.21 and 0.18, respectively.

The grid blocks in the SC domain consist of $6 \times 8 \times 8$ cells. To improve the spatial resolution of the simulation, the adaptive mesh refinement (AMR) library \citep{Toth2012} is applied to refine the grid in the lower corona ($1.05 - 6.0$~{\rsun}) and regions near the heliospheric current sheet (HCS). As a result, the total number of grid cells in the SC domain reaches 17.1 million. With AMR, the angular resolution of the grid below 6.0~{\rsun} improves from an initial value of 1.4{\dg} to 0.7{\dg}, corresponding to approximately 8.54 Mm on the solar surface and $11.78''$ as seen from the Earth. The radial resolution near the solar disk is about $2 \times 10^{-4}$~{\rsun} (150 km). The simulation starts at 19:30 UT on 2024 April 8. The SC component iterates for a total of 80,000 steps to reach the numerical steady-state solution of the MHD equations.

\section{Results} \label{sec3}

The MHD simulations of the 2024 TSE are conducted using the AWSoM-R model applying the free parameters and boundary conditions described in Section \ref{sec2}. Simulation results obtained by different synoptic maps, as shown in Figure \ref{fig2}, are compared aiming to investigate the influence of far-side magnetic structures on the large-scale coronal configuration.

\subsection{Simulated Plasma Parameters}

\begin{figure*}[htb!]
\begin{center}
\includegraphics[width=0.9\textwidth]{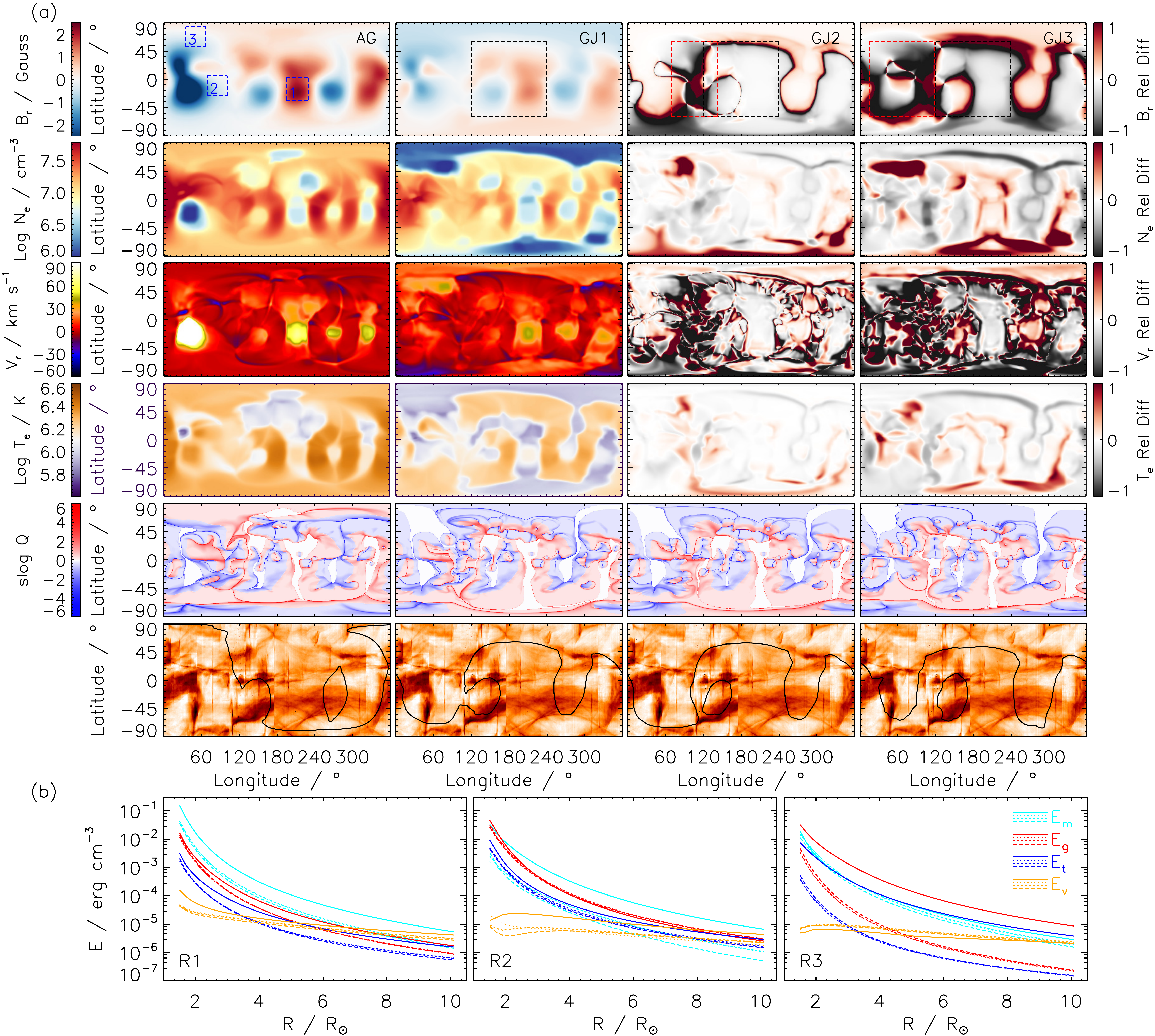}
\caption{Comparisons of plasma parameters and energy density derived from AWSoM-R model using different synoptic maps. (a) Plasma parameter distributions of the radial magnetic field (first row), electron number density (second row), radial speed (third row), and electron temperature (fourth row) on the spherical shell at a heliocentric distance of 1.5~{\rsun}. The left to right columns present results of the AG and GJ1–GJ3 maps, respectively. The GJ2 and GJ3 results, indicated by the red-black colorbar, represent the relative differences compared to the GJ1 distributions. The black box represents the region where the full-disk magnetic field on that day is replaced in the original GJ1 map. The red boxes represent the regions where the full-disk GONG magnetograms from 4 and 8 days later are replaced, respectively. The fifth row presents the signed log Q distribution corresponding to closed magnetic fields at the photospheric boundary. The sixth row presents the comparison between the coronal Carrington map and the simulated HCS at 3.0~{\rsun}. (b) Energy density profiles as a function of heliocentric distance for the regions marked by blue boxes in panel (a), including the active region (R1), the quiet region (R2), and the polar region (R3). \label{fig3}}
\end{center}
\end{figure*}

Figure \ref{fig3} (a) presents longitude-latitude diagrams of various parameters, including the radial magnetic field $B_r$, electron number density $N_e$, radial speed $V_r$, electron temperature $T_e$, quasi-separatrix layer (QSL), and HCS distribution, from top to bottom rows. The columns, from left to right, correspond to simulation results based on the AG and GJ1–GJ3 maps. The distributions of plasma parameters are extracted on a spherical shell at a heliocentric distance of 1.5~{\rsun} within the SC domain. Since it is difficult to visually compare the detailed features among the results of the GJ1-GJ3 maps, the parameter distributions M of the GJ2 and GJ3 maps indicated by the red-black colorbar in panel (a) are relative differences compared to the GJ1 map, defined as ${\rm M~Rel~Diff} = {\rm M}/{\rm M}_{\rm GJ1} - 1$. The black dashed box represents the region where the original GJ1 map is replaced with the same-day magnetogram, while the red dashed boxes indicate regions where full-disk magnetic field observations after 4 and 8 days are replaced in this work. The blue dashed boxes indicate the regions where energy density profiles are extracted, including the active region (R1), quiet region (R2), and polar region (R3). Panel (b) presents the corresponding results as a function of heliocentric distance. The simulation results indicate that plasma parameter values derived from the AG map are generally higher than those obtained by the GJ1-GJ3 maps. This discrepancy may be attributed to the stronger positive magnetic flux in the AG map, as well as the corrections applied to the polar and weak-field regions using the flux transport model. Compared to the GJ1 map, the results from the GJ2 and GJ3 maps present significant changes in the distributions of magnetic field and speed within the replacement regions (red dashed boxes). The influence of far-side magnetic structures on the global plasma parameter distribution is primarily concentrated near the HCS, with noticeable effects even in the polar regions. This effect increases with extending the time interval used to replace the magnetogram in the synoptic map and is particularly evident in the results of the GJ3 map, which incorporates full-disk magnetic field data from 8 days later.

The fifth row in Figure \ref{fig3} (a) presents the distributions of the signed log Q, defined as ${\rm slog}\,Q = {\rm sgn}(B_{\rm z})\cdot {\rm log}\,Q$, corresponding to the closed field lines at the photosphere calculated by the improved FastQSL code (\citealt{Zhang2022ApJ}; J. Chen et al. 2025, in prep.) in the spherical coordinate to quantify the connectivity of the magnetic field, while open magnetic fields are shown as white regions. Compared to the GJ1-GJ3 maps, the AG map significantly improves the simulation of the magnetic field in the polar regions, reducing the open magnetic field area to 8.81\% of the total surface. This further supports the influence of polar magnetic fields on the open flux, as discussed by \citet{Riley2019} and \citet{Shi2024}. In contrast, simulations using the GJ1, GJ2, and GJ3 maps yield open field areas of 14.22\%, 14.13\%, and 14.90\%, respectively. These results indicate that the correction of the local magnetic field on the eastern solar hemisphere not only affects simulation results within this specific active region but also changes the distribution, connectivity, and open field regions of the global magnetic field. The bottom row of panel (a) presents the comparisons of the coronal Carrington map composed of slices extracted from the LASCO C2 images at a heliocentric distance of 3.0~{\rsun} during CR 2282 with the simulated HCS (black line). The Carrington map is displayed using a reverse color table to enhance coronal structures, where dark regions correspond to brighter streamers. The HCS is determined based on the magnetic neutral line on the spherical shell of 3.0~{\rsun} in the SC domain. In the AG simulation, the HCS extends up to nearly 90{\dg} latitude, showing a significant deviation from the structure observed in the Carrington map. Compared to the results from the GJ1 and GJ2 maps in the longitude ranges of 40{\dg}–100{\dg} and 250{\dg}–300{\dg}, the GJ3 simulation shows discrepancies in the coronal structures at mid-to-high southern latitudes relative to observations. The GJ2 result exhibits good agreement with observations except in the longitude range of 0–48{\dg}. These results confirm that the local far-side magnetic field influences the distribution of the initial coronal magnetic field extrapolated by the PFSS model, which further alters the simulation results after iterating the MHD equations. The synoptic map spliced from a single Earth-based view, even after applying the flux transport model correction, cannot accurately reproduce the location of the HCS. Furthermore, correcting the magnetic field in the observationally blind regions of the synoptic map is more effective for generating a reliable simulated HCS.

Figure \ref{fig3} (b) presents the energy density profiles in the active region (R1), quiet region (R2), and polar region (R3) from the MHD simulations. The cyan, red, blue, and orange lines represent the profiles of magnetic energy density, gravitational potential energy density, thermal energy density, and kinetic energy density, respectively, as a function of the heliocentric distance $r$ (1.5-10.1~{\rsun}), calculated using the equations in Appendix \ref{sec:apd2}. The solid, dotted, short-dashed, and long-dashed lines correspond to simulations using the AG map and the GJ1–GJ3 maps, respectively. The energy density evolution in the GJ1–GJ3 simulations exhibits a similar trend. In the active region, the AG simulation consistently yields higher energy densities than the GJ1–GJ3 maps. Magnetic energy dominates, significantly exceeding gravitational potential and thermal energy, while kinetic energy gradually approaches or even surpasses magnetic energy with increasing $r$. In the quiet region, the magnetic and kinetic energy densities simulated by the AG map are significantly higher than those from the GJ1-GJ3 maps, while the gravitational potential and thermal energy densities remain similar. The kinetic energy derived from the AG map initially increases and then decreases with $r$, exhibiting an inverse trend compared to the GJ1–GJ3 simulations. In the AG case, magnetic energy is dominant, whereas in the GJ1–GJ3 cases, gravitational potential energy prevails. In the north polar region, the simulation based on the AG map yields significantly higher gravitational potential and thermal energy densities compared to the GJ1–GJ3 maps, while the magnetic energy density is slightly higher and the kinetic energy density is slightly lower. Here, gravitational potential energy dominates in the AG simulation, whereas magnetic energy is dominant in the GJ1–GJ3 cases. These results confirm that the relationship between magnetic energy and gravitational potential energy does not strictly satisfy the empirical energy equipartition in the corona, as derived by \citet{Bemporad2023ApJ}, and may require more reliable observations for further constraints. Additionally, the coronal energy density depends strongly on the choice of synoptic map.

\begin{figure*}[htb!]
\begin{center}
\includegraphics[width=0.8\textwidth]{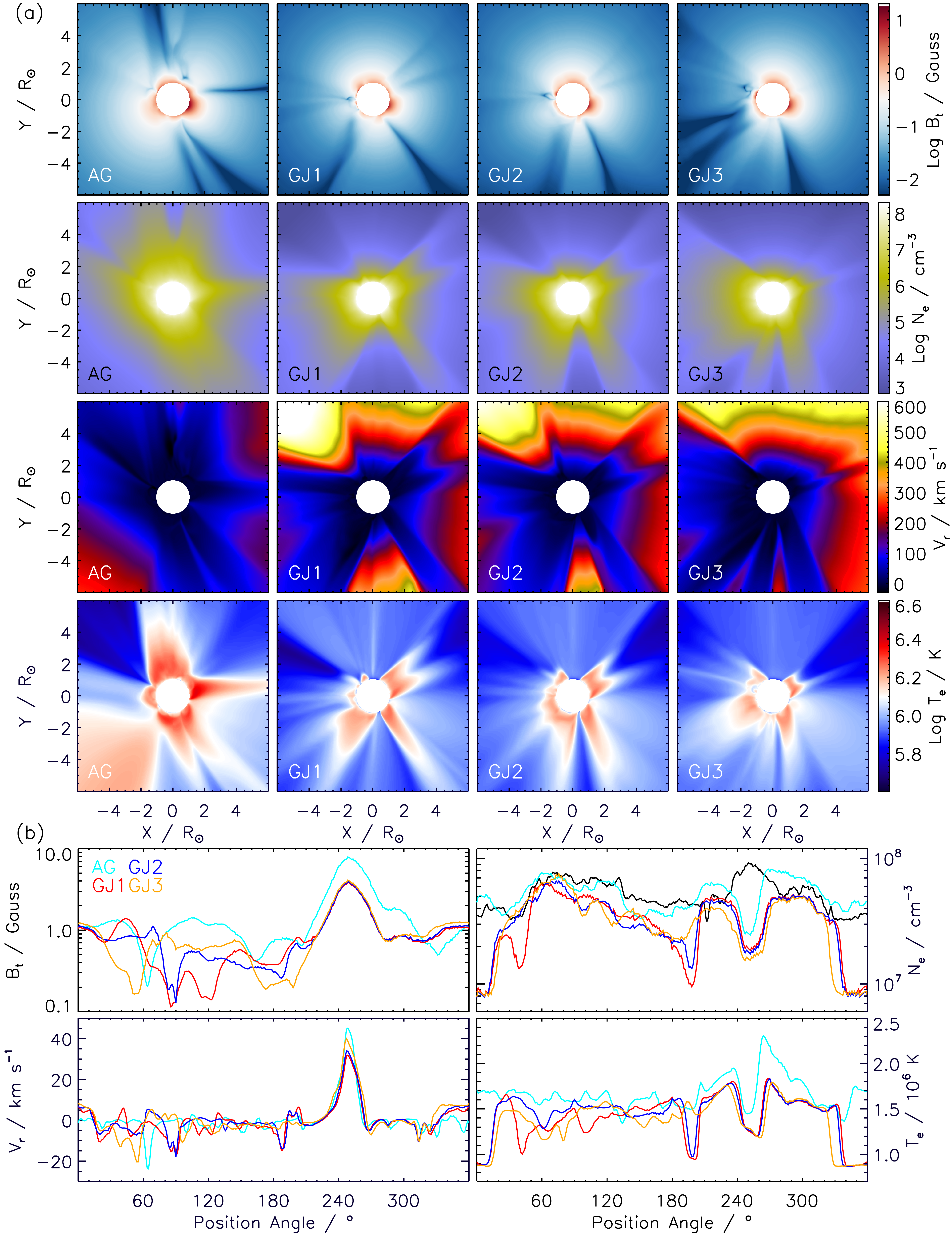}
\caption{Comparison of simulated plasma parameters on the PoS. (a) Distributions of plasma parameters in the low corona ($-6.0 \sim 6.0$~{\rsun}) as viewed from Earth. The top to bottom rows represent the total magnetic field strength, electron number density, radial speed, and electron temperature. The columns from left to right correspond to simulation results using the AG and GJ1-GJ3 maps as boundary conditions. (b) Plasma parameter profiles extracted from the heliocentric distance of 1.2~{\rsun} on the PoS. The colored lines represent the simulation results, while the black line represents the electron number density profile inverted from the coronal pB observed during the 2024 TSE. \label{fig4}}
\end{center}
\end{figure*}

The distribution of plasma parameters on the Plane of Sky (PoS) as viewed from Earth is presented in Figure \ref{fig4} (a). From the top to bottom rows, the total magnetic field strength $B_t$, electron number density $N_e$, radial speed $V_r$, and electron temperature $T_e$ are presented in the low coronal region of $-6.0 \sim 6.0$~{\rsun}. The southeastern coronal region simulated by the AG map has higher solar wind speed and electron temperature, showing a significant contrast with the results derived from the GJ1-GJ3 maps. These differences arise from modifications to the original distribution and strength of the photospheric magnetic field after applying the flux transport model to the AG map. A comparison of the GJ1–GJ3 maps reveals that far-side magnetic structures influence the deflection direction of streamers, causing them to incline more toward the north and south poles, thereby influencing the global distribution of coronal structures. The simulation using the GJ3 map presents streamer structures on the eastern limb characterized by a weaker magnetic field and solar wind speed, along with higher electron density and temperature, indicating that the far-side photospheric active region significantly impacts the simulation results of the corresponding upper corona. Panel (b) presents a quantitative comparison of the plasma parameter profiles extracted at a heliocentric distance of 1.2~{\rsun} on the PoS as a function of position angle (PA). The cyan, red, blue, and orange lines correspond to simulation results of the AG, GJ1-GJ3 maps, respectively. The black line represents the electron number density profile derived from the TSE pB observations using the inversion method proposed by \citet{Hulst1950}, which assumes an axisymmetric density distribution expressible in polynomial form. The density can be obtained by fitting the pB profile along the radial direction. In the low corona region, except for the radial speed, the plasma parameters simulated using the AG map are significantly higher than those derived from the GJ1–GJ3 maps. The electron number density obtained from the AG map shows better agreement with the observed inversion profile, while the results of the GJ1-GJ3 maps are underestimated. The underestimation may originate from the relatively weaker magnetic field strength measured in the original GONG full-disk magnetogram. Notably, the discrepancy between the observed and simulated densities in the PA range of 230{\dg} $\sim$ 270{\dg} may be attributed to the line-of-sight (LoS) integration effect inherent in the observation, which is not explicitly included in the model-extracted PoS profile.

\begin{figure*}[htb!]
\begin{center}
\includegraphics[width=0.9\textwidth]{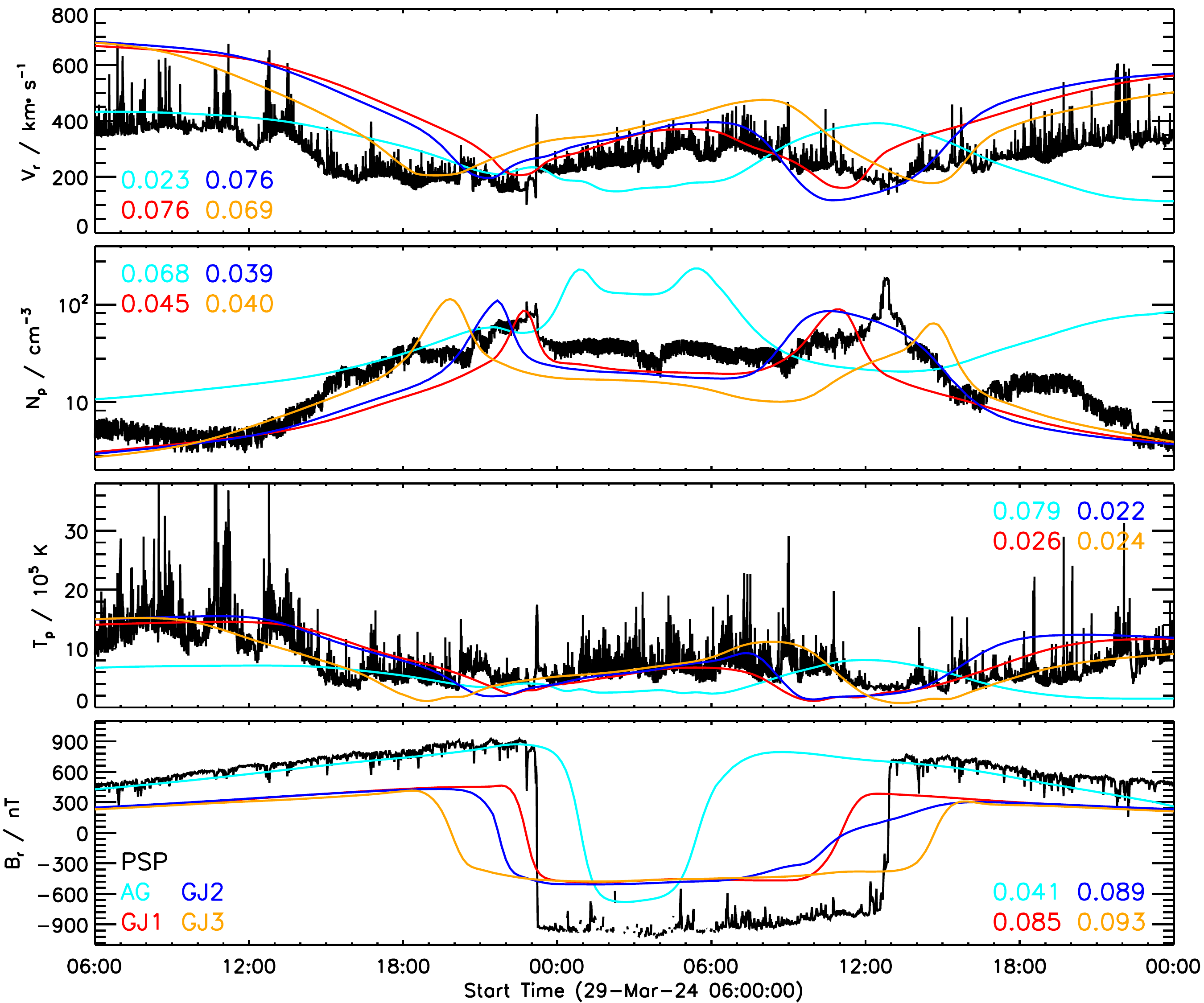}
\caption{Comparison of plasma parameters measured in situ by PSP (black line) with simulation results based on the AG and GJ1-GJ3 maps (colored lines). The Dist index marked in the corners quantifies the agreement between the observations and the model predictions. \label{fig5}}
\end{center}
\end{figure*}

\subsection{Comparisons with In Situ Measurements}

To further validate the model results, we compared the simulated plasma parameters with in situ measurements from PSP during its nineteenth solar encounter (E19), spanning from 06:00 UT on March 29 to 00:00 UT on March 31, 2024. During this period, the PSP reached its perihelion of 11.4~{\rsun} on March 30. The plasma parameters and magnetic field represented by the black line in Figure \ref{fig5} are measured by the Solar Wind Electrons Alphas and Protons \citep[SWEAP;][]{Kasper2016} and Electromagnetic Fields Investigation \citep[FIELDS;][]{Bale2016} instruments, respectively. Specifically, the radial speed $V_r$, proton number density $N_p$, and proton temperature $T_p$, shown from the first to the third row, are collected by the Solar Probe ANalyzers for Ions \citep[SPAN-I;][]{Livi2022}, which is one of the instruments in the SWEAP instrument suite. The colored lines represent the plasma parameters extracted along the PSP orbit within the SC domain during this time period. The Dist index proposed by \citet{Sachdeva2019} is shown in the corner of each panel to quantitatively characterize the error between the observations and model predictions. It is defined as,
\begin{equation}
{\rm Dist} = \frac {D_{1, 2}+D_{2, 1}}{2},
\end{equation}
where $D_{1, 2}$ ($D_{2, 1}$) is the average of the minimum distances between PSP measurements (curve 1) and the simulation results (curve 2) integrated along curve 1 (2). The smaller the Dist value, the more accurate the simulation results. The comparison shows that the radial speed and magnetic field simulated using the AG map (cyan line) are more consistent with the measurements than those using the GJ1–GJ3 maps, while the GJ2 map (blue line) provides better results for proton density and temperature. This indicates that the AG map, after correction by the flux transport model, exhibits stronger magnetic field strength, resulting in lower simulated solar wind speed and temperature, but higher density. The radial magnetic fields simulated using the GJ1-GJ3 maps are underestimated compared to the PSP measurements, called the Open Flux Problem \citep[OFP, ][]{Linker2017}, which implies that the OFP is scarcely influenced by the far-side magnetic field. In contrast, results from the AG map indicate that the polar magnetic field plays an important role in mitigating the OFP \citep[e.g.][]{Riley2019, Shi2024}. Nevertheless, the overall evolution trends of the simulations based on the GJ1–GJ3 maps remain consistent, with noticeable time shifts between the peak and trough values. In situ comparisons of plasma parameters further confirm that replacing the far-side magnetic field in the synoptic map can indeed improve the accuracy of the coronal modeling.

\begin{figure*}[htb!]
\begin{center}
\includegraphics[width=0.9\textwidth]{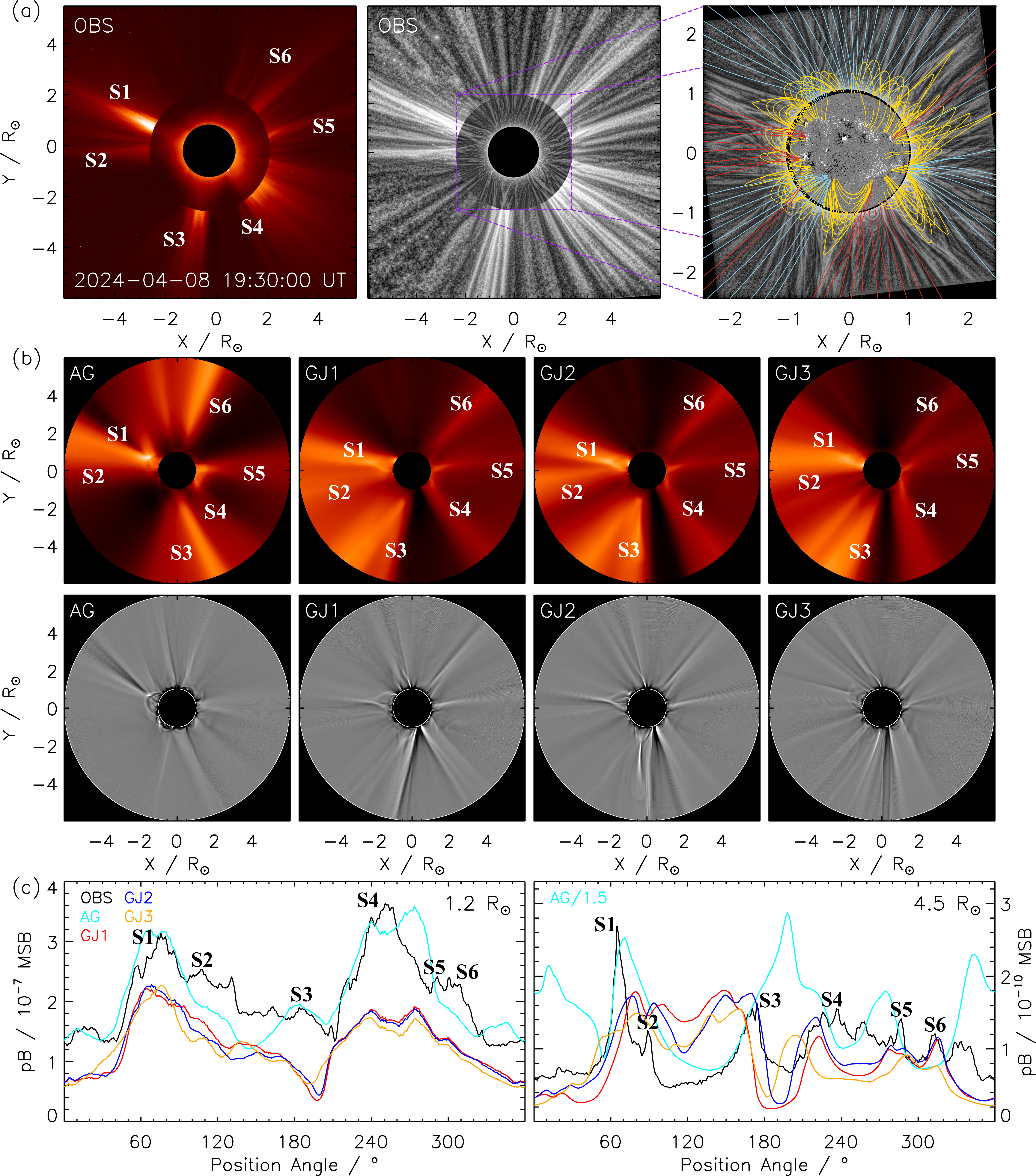}
\caption{Comparison of simulated and observed WL images within the FoV of TSE and LASCO C2. (a) Composite observations of TSE and LASCO C2, and an image enhanced using the WOW algorithm. The zoomed-in region, indicated by the purple box, is overplotted with the magnetic field lines traced from the simulation results using the GJ3 map. (b) The first row is WL synthetic images generated from simulations using the AG and GJ1-GJ3 maps, respectively, with enhancement applied using the NRGF algorithm. The second row presents difference images between the synthetic WL images and their corresponding smoothed versions to enhance fine coronal structures. (c) Comparison between the simulated pB profiles (colored lines) and the radiometrically calibrated observations (black lines) from the 2024 TSE (left) and LASCO C2 (right) at heliocentric distances of 1.2~{\rsun} and 4.5~{\rsun}, respectively. \label{fig6}}
\end{center}
\end{figure*}

\subsection{Comparisons with WL Observations}

Based on the Thomson scattering mechanism of free electrons in the corona, the intensity distribution of WL radiation can be calculated from the simulated electron number density, and the synthesized WL image is obtained by integrating it along the LoS. Figure \ref{fig6} presents a comparison between the observed and simulated WL images of the 2024 TSE. Panel (a) shows the composite image of the LASCO C2 observation subtracted by the monthly-minimum background and the radially filtered 2024 TSE observation (first column) to enhance the contrast of streamers, as well as the image enhanced using the WOW algorithm (second column), in order to highlight fine coronal structures. The purple box indicates a zoomed-in region of the TSE image (third column), overplotted with coronal magnetic field lines traced from the simulation results using the GJ3 map. The red and blue lines denote open magnetic field lines with positive and negative polarity, respectively, while the yellow lines represent closed magnetic field lines. The simulated magnetic field effectively reproduces the true coronal structures. The orientations of the open field lines are basically in alignment with the plumes, while the positions of the closed field lines correspond well to the streamers. Panel (b) presents the synthesized WL images processed by the normalizing radial gradient filter (NRGF) algorithm (first row), as well as the difference images between the original WL simulated images and their smoothed images (second row), further highlighting fine structures in the simulated corona. The corresponding synoptic map is marked in the corners. It can be seen that the large-scale coronal structures simulated using the GJ1–GJ3 maps reproduce the TSE observations very well compared with those from the AG map, although slight differences exist in the streamer orientations. Panel (c) presents a quantitative comparison of the pB as a function of PA, extracted from the 2024 TSE and LASCO C2 observations at heliocentric distances of 1.2~{\rsun} (left column) and 4.5~{\rsun} (right column), along with the corresponding simulated images. The black lines represent the observed pB derived from the radiometrically calibrated TSE and the LASCO C2 Legacy Archive\footnote{\url{http://idoc-lasco.ias.u-psud.fr/}}, while the colored lines correspond to the simulation results using the AG, GJ1-GJ3 maps, respectively.

The evaluation of the simulation quality focuses on comparing the deflection directions and radiation intensities of the 6 prominent streamers labeled as ``S1''-``S6'' in Figure \ref{fig6} during the 2024 TSE. As the most commonly applied map in global MHD simulations, the pB profile at a heliocentric distance of 1.2~{\rsun} simulated using the AG map shows good agreement with the TSE observation in magnitude, especially for the ``S1'' and ``S3'' streamers. The pB profiles derived from the GJ1-GJ3 maps exhibit overall variation trends more consistent with the observations than the AG map; however, they underestimate the pB by approximately 39.8\%. The right column of panel (C) indicates that the pB profiles derived from the GJ1–GJ3 maps at 4.5~{\rsun} are in good agreement with the LASCO C2 observation. In contrast, the AG map result requires a scaling factor of 1.5 for better visualization and comparison. Although the pB can be brought into agreement with the observations by adjusting the simulation parameter $(S_{\rm A}/B)_\odot$, such modification would result in synthesized coronal structures, as shown in panel (b), that deviate significantly from the observed morphology. The direction of the ``S1'' streamer simulated by the AG map is accurately reproduced in agreement with the observations, while the ``S2''-``S6'' streamers exhibit varying degrees of deviation. Notably, the ``S4'' streamer simulated by the AG map has a relatively low cusp height and does not extend into the high corona as shown in the observation. Additionally, the AG simulation results have large peaks in the PA ranges of 0 $\sim$ 50{\dg}, 180{\dg} $\sim$ 210{\dg}, and 320{\dg} $\sim$ 360{\dg}, opposite to both the LASCO C2 observation and the simulation results of the GJ1-GJ3 maps. These differences may be attributed to the flux transport model applied to calibrate the AG map, which alters the original magnetic flux and consequently introduces systematic biases that affect the simulated large-scale structure of the corona. A comparison between the simulations based on the GJ1–GJ3 maps and the pB profiles derived from LASCO C2 reveals that replacing the far-side magnetic field significantly influences the streamer directions on both the eastern and western limbs. This effect may arise from the north–south magnetic pressure imbalance associated with the polar magnetic field variations shown in Figure \ref{fig3} (a). Specifically, except for the ``S4'' streamer, which exhibits southward deflections of 1.2{\dg}-18.0{\dg} at 4.5~{\rsun}, and the direction of ``S6'' streamer remains nearly unchanged, almost all other streamers exhibit northward deflections with an average of 3.6{\dg} for GJ2 map and 16.8{\dg} for GJ3 map compare to GJ1 results. Regarding radiation intensity, the pB gradually decreases by 18.4\% from GJ1 to GJ3, with a slight increase of 10.2\% on the western limb in the GJ2 simulation. Interestingly, the GJ2 simulation exhibits two peaks at the ``S3'' streamer position, with the direction of the second peak consistent well with the observed structure, potentially related to variations in the polar field strength and the HCS position. Overall, the GJ2 map, which replaces the full-disk magnetogram 4 days later, provides the best agreement with observations in terms of streamer deflection direction. This suggests that extending the time interval for replacing the magnetogram does not necessarily improve simulation accuracy, as the far-side magnetic field undergoes dynamic changes over time. Meanwhile, the GJ3 results highlight the crucial role of the far-side active region on the eastern hemisphere in calibrating streamer deflection at the western limb.

\section{Summary} \label{sec4}

The corona is a critical region connecting the solar surface to the solar wind, with its magnetic field structure and plasma dynamics directly determining the physical properties of the solar wind. Additionally, the corona is the main region of solar activity, where eruptive events such as flares and CMEs originate, significantly impacting Earth and its surrounding space environment. Accurate modeling of the corona is therefore of great scientific and practical importance. Global MHD simulations rely on real-time, high-precision photospheric magnetic field measurements. However, the routine synoptic map, based on Earth-view observations, is inherently limited in capturing far-side magnetic structures. To address this limitation, this work corrects the far-side active regions in the GONG Janus map by incorporating full-disk GONG magnetograms measured a few days after the 2024 TSE, and employs them as boundary conditions for the AWSoM model to investigate their impact on the simulation results.

Our research find that the synoptic map after replacing the magnetic field of the far-side active region alters the original positive-negative flux balance of the photosphere, which not only affects the simulated plasma parameters in the local upper corona but also influences their distribution across the entire corona, particularly near the HCS and the polar regions. Significant discrepancies arise in the distribution of coronal energy densities and plasma parameters simulated using the AG map compared with the GJ1–GJ3 maps. Comparisons with observations from the PSP and the 2024 TSE + LASCO C2 further demonstrate that far-side field corrections in the synoptic map can substantially improve the accuracy of global coronal modeling, but they cannot fundamentally resolve the OFP. Specifically, the GJ2 map constructed by replacing the far-side region in the original GJ1 daily-updated synoptic map with the full-disk magnetogram observed 4 days after the TSE yields simulated proton number density, proton temperature, and streamer deflection direction that are relatively consistent with observations. The most notable change is the streamers on the eastern and western limbs of the synthetic WL image deflect toward the poles, which may result from the shift of the HCS position caused by the polar magnetic pressure.

The approach for correcting the synoptic map assumes that the magnetic structures of the far-side active region remain relatively stable in the next few days, partially mitigating the observational blind spots caused by the limitations of current single-view magnetic field measurements, which can reflect the evolution of the photospheric magnetic field more accurately and in real time, and provide more reliable data for model inputs. It is worth noting that the current GONG Janus map underestimates the magnetic field strength compared to the AG map, leading to a lower simulated pB relative to the observations. Although the AG map corrected by the flux transport model can reproduce the magnitude of the pB in the lower corona, the simulated streamer orientations deviate substantially from the LASCO C2 observation. Our findings provide new insights into the relationship between the coronal structure and the magnetic field evolution, emphasizing the necessity of developing multi-view magnetic field measurement technology. Future research endeavors will combine magnetic field measurements from SolO/PHI and the Solar Polar-orbit Observatory \citep[SPO;][]{Deng2025} to perform real-time corrections of far-side and polar regions in synoptic maps, thereby enhancing the accuracy of global MHD simulations. Additionally, we will further investigate the impact of different resolutions of the bottom boundary conditions on coronal simulation results.

\begin{acknowledgments}

Guanglu Shi and Jiahui Shan contributed equally to this manuscript and are co-first authors. We sincerely thank the anonymous referee for providing valuable suggestions that helped us improve the quality of the manuscript. We thank Hongxin Zhang for assistance with the TSE observation and data calibration, and Jinsong Zhao for helpful discussions regarding the processing of PSP data. The authors thank the GONG and SOHO/LASCO teams for their open-data use policy. GONG is a community-based program, managed by the National Solar Observatory (NSO). SOHO is an international cooperative mission between ESA and NASA. This work makes use of the LASCO C2 legacy archive data produced by the LASCO C2 team at the Laboratoire d'Astrophysique de Marseille and the Laboratoire Atmosph\`{e}res, Milieux, Observations Spatiales, both funded by the Centre National d'Etudes Spatiales (CNES). This work utilizes the ADAPT-GONG map produced collaboratively between AFRL/ADAPT and NSO/NISP. This work is supported by the Strategic Priority Research Program of the Chinese Academy of Sciences, Grant No. XDB0560000, National Key R\&D Program of China 2022YFF0503003 (2022YFF0503000), the project of Solar Polar Observatory GJ11020204, GJ11020403 and GJ11020405, NSFC (grant Nos.12233012, 12203102). This work benefits from the discussions of the ISSI-BJ Team ``Solar eruptions: preparing for the next generation multi-waveband coronagraphs''.

\end{acknowledgments}

% \begin{contribution}

% %This section gives authors the space to recognize author contributions. The text inside this environment is NOT counted towards the total word quanta. At a minimum, manuscripts are expected to include this text:

% All authors contributed equally to the Terra Mater collaboration.

% % But authors are expected to provide more specific details, e.g. 
% %
% %SC was responsible for writing and submitting the manuscript.
% %WWM came up with the initial research concept and edited the manuscript.
% %OTS obtained the funding and edited the manuscript.
% %EBF provided the formal analysis and validation. He also edited the manuscript.
% %GEH Supervised the undergraduates, wrote the software and administers the project github and Zenodo repositories.
% %
% % Authors can use the Contributor Role Taxonomy (CRediT) at
% % https://credit.niso.org
% % for ideas on how write a good statement tailored to their needs.

% \end{contribution}

% \facilities{HST(STIS), Swift(XRT and UVOT), AAVSO, CTIO:1.3m, CTIO:1.5m, CXO}


% \software{astropy \citep{2013A&A...558A..33A,2018AJ....156..123A,2022ApJ...935..167A},  
%           Cloudy \citep{2013RMxAA..49..137F}, 
%           Source Extractor \citep{1996A&AS..117..393B}
%           }

\appendix

\section{AWSoM Model} \label{sec:apd1}

The AWSoM is a physics-based model integrated into the SWMF, utilizing the SC and Inner Heliosphere (IH) components to perform 3D MHD simulations of the solar wind from the corona into interplanetary space. AWSoM self-consistently describes key physical processes such as coronal heating and solar wind acceleration by modeling {\alfven} wave interactions, which nonlinearly couples both forward and backward propagating low-frequency {\alfven} waves along magnetic field lines, leading to turbulence cascade and dissipative heating \citep{Velli1989, Zank1996, Matthaeus1999, Chandran2011, Zank2012}. The model considers the collision heat conduction of electrons and the radiation loss solved by the CHIANTI database \citep{Dere1997}, thereby separating the temperatures of electrons and protons \citep{Sokolov2013ApJ}. Additionally, by introducing a more self-consistent treatment of {\alfven} wave reflection and stochastic heating, AWSoM effectively resolves the anisotropic proton temperature (parallel and perpendicular components) and the associated kinetic instabilities in the solar wind \citep{Holst2014ApJ, Meng2015, Holst2019, Holst2022}. The AWSoM-R model \citep{Sokolov2021} further optimizes the treatment of the low corona by assuming a nearly steady-state, low-$\beta$ plasma with heat and mass transport along magnetic field lines, reducing computational costs and execution time for near real-time solar wind and CME predictions.

The 2024 TSE is simulated using the AWSoM-R model, which uses the numerical Block-Adaptive-Tree-Solarwind-Roe-Upwind-Scheme (BATS-R-US) to solve the following MHD governing equations,
\begin{equation}
	\frac{\partial \rho}{\partial t} + \nabla \cdot (\rho \bm{u}) = 0,
\end{equation}
\begin{equation}
	\frac{\partial \bm{B}}{\partial t} - \nabla \times (\bm{u} \times \bm{B}) = 0,
\end{equation}
\vspace{-10pt}
\begin{align}
	\frac{\partial (\rho \bm{u})}{\partial t} + \nabla & \cdot \left(\rho\bm{uu} - \frac{\bm{BB}}{\mu_0} \right) + \nabla \left(P_i + P_e \right. \nonumber \\
    \left. \right. & \left. + \frac{B^2}{2\mu_0} + P_A \right) = - \rho \frac{GM_\odot}{r^3}\bm{r},
\end{align}
\vspace{-10pt}
\begin{align}
	\frac{\partial}{\partial t} \left(\frac{P_i}{\gamma - 1} \right. & \left. + \frac{\rho \mu^2}{2} + \frac{B^2}{2\mu_0} \right) + \nabla \cdot \left[\left(\frac{\rho u^2}{2} \right. \right. \nonumber \\
    \left. \left. + \right. \right. & \left. \left. \frac{\gamma P_i}{\gamma - 1} + \frac{B^2}{\mu_0} \right)\bm{u} - \frac{\bm{B}(\bm{u} \cdot \bm{B})}{\mu_0}\right] \nonumber \\
    = - \bm{u} \cdot \nabla & (P_e + P_A) + \frac{N_eN_ik_B}{\gamma - 1} \left(\frac{\nu_{ei}}{N_i}\right)(T_e - T_i) \nonumber \\
	+ f_p(&\Gamma_-w_-+\Gamma_+w_+) - \rho \frac{GM_\odot}{r^3} \bm{r} \cdot \bm{u},
\end{align}
\vspace{-10pt}
\begin{align}
	\frac{\partial}{\partial t} \left(\right. & \left. \frac{P_e}{\gamma - 1} \right) + \nabla \cdot \left(\frac{P_e}{\gamma - 1}\bm{u} \right) + P_e \nabla \cdot \bm{u} \nonumber \\
	= & - \nabla \cdot (\kappa \cdot \nabla T_e) + \frac{N_eN_ik_B}{\gamma - 1}\left(\frac{\nu_{ei}}{N_i}\right)(T_i - T_e) \nonumber \\
	& - Q_{\rm rad} + (1 - f_p)(\Gamma_-w_-+\Gamma_+w_+),
\end{align}
where $\rho$ is the mass density of the plasma, $\bm{u}$ is the plasma fluid speed, and $\bm{B}$ is the magnetic field. The parameter $\mu_0$ is the vacuum permeability, $G$ is the gravitational constant, $M_\odot$ is the solar mass, while $\bm{r}$ represents the heliocentric distance. The proton and electron number densities are given by $N_{i, e}=\rho/m_p$, where $m_p$ is the proton mass. The temperatures of protons and electrons are represented by $T_{i, e}$, and $k_B$ is the Boltzmann constant. The thermal pressures of protons and electrons are defined as $P_{i, e} = N_{i, e}k_BT_{i, e}$. The parameters $w_+$ and $w_-$ represent the energy densities of {\alfven} waves propagating along and against the magnetic field lines, respectively, with the {\alfven} wave pressure given by $P_A = (w_+ + w_-)/2$. The terms $\Gamma_{+, -}$ are the dissipation rates of {\alfven} waves. $\nu_{ei}$ is the Coulomb collision frequency between electrons and protons, while $Q_{\rm rad}$ describes the radiative energy loss function for optically thin plasma. The ion heating fraction, $f_p$, quantifies the partitioning of energy between ions and electrons. The parameter $\kappa$ is the thermal conductivity coefficient, and the adiabatic index $\gamma$ is assumed to be $5/3$.

\section{Energy Density} \label{sec:apd2}

The distribution of energy density in different coronal regions can be derived from the AWSoM simulation results. The magnetic energy density $E_m$, gravitational potential energy density $E_g$, thermal energy density $E_t$, and kinetic energy density $E_v$ are expressed as functions of the heliocentric distance $r$ as follows,
\begin{align}
	& E_m = B^2(r)/2\mu_0,  \\
	& E_g = -GM_\odot \rho (r)/r, \\
	& E_t = 2N_e(r) k_B T_e(r), \\
	& E_v = \rho(r)V(r)^2/2,
\end{align}
where the magnetic field $B$, mass density $\rho$, electron number density $N_e$, electron temperature $T_e$, and speed $V$ are all derived from the simulation.

\bibliography{shigl_biblio}{}
\bibliographystyle{aasjournal}



\end{document}